\documentstyle[aps,amsfonts,eqsecnum]{revtex}
\begin{document}
\draft
\title{The Bargmann representation for the quantum mechanics on a
sphere}
\author{K. Kowalski and  J. Rembieli\'nski}
\address{Department of Theoretical Physics, University
of \L\'od\'z, ul.\ Pomorska 149/153,\\ 90-236 \L\'od\'z,
Poland}
\maketitle
\begin{abstract}
The Bargmann representation is constructed corresponding to the
coherent states for a particle on a sphere introduced in:
K. Kowalski and J. Rembieli\'nski, J. Phys. A: Math. Gen. {\bf 33}, 
6035 (2000).  The connection is discussed between the introduced 
formalism and the standard approach based on the Hilbert space of 
square integrable functions on a sphere $S^2$.
\end{abstract}
\section{Introduction}
In our recent paper \cite{1} the coherent states for a particle on a
sphere have been introduced.  As with the standard coherent states
\cite{2} those states are labelled by points of the classical phase space
i.e.\ the cotangent bundle $T^*S^2$.  It is worthwhile to recall
that the celebrated spin coherent states \cite{3} are not related 
to the phase space for a particle on the sphere $S^2$.  One of
the most characteristic properties of coherent states is the
existence of the Bargmann representation.  Such representation is of
importance not only from the mathematical point of view.  An example
of applications are the Husimi functions i.e.\ the elements of the
Bargmann space, in the theory of quantum chaos.  In this work we
introduce the Bargmann representation referring to the coherent
states for a particle on a sphere mentioned above.  It should be noted 
that, in opposition to the case of the standard coherent states, the
construction of such Bargmann representation is a highly nontrivial problem.  
The paper is organized as follows.  In section II we recall the basic 
properties of the coherent states for a particle on a sphere.  Sections III--V
are devoted to the construction of the Bargmann representation.  In section VI 
we discuss the connection of the introduced Bargmann representation and the 
standard coordinate representation for the quantum mechanics on a sphere.
\section{Coherent states for a particle on a sphere}
Our purpose in this section is to recall the basic properties of the
coherent states for a particle on a sphere introduced in \cite{1}.
Those states are related to the $e(3)$ algebra of the form
\begin{equation}
[J_i,J_j]={\rm i}\varepsilon_{ijk}J_k,\qquad [J_i,X_j]={\rm i}
\varepsilon_{ijk}X_k,\qquad [X_i,X_j]=0,\qquad i,\,j,\,k=1,\,2,\,3.
\end{equation}
The Casimir operators are given in a unitary irreducible representation by
\begin{equation}
{\bf X}^2=r^2,\qquad {\bf J}\bbox{\cdot}{\bf X}=\lambda,
\end{equation}
where dot designates the scalar product.  In \cite{1} we restricted
to the special case $\lambda=0$, so
\begin{equation}
{\bf J}\bbox{\cdot}{\bf X}=0.
\end{equation}
The irreducible representation of (2.1) under the choice (2.3) is spanned
by the common eigenvectors $|j,m;r\rangle$ of the operators ${\bf J}^2$,
${\bf X}^2$ and ${\bf J}\bbox{\cdot}{\bf X}$.  We have
\begin{mathletters}
\begin{eqnarray}
&&{\bf J}^2 |j,m;r\rangle = j(j+1) |j,m;r\rangle,\qquad J_3
|j,m;r\rangle=m|j,m;r\rangle,\\
&&{\bf X}^2 |j,m;r\rangle=r^2 |j,m;r\rangle,\qquad
({\bf J}\bbox{\cdot}{\bf X}/r) |j,m;r\rangle=0,
\end{eqnarray}
\end{mathletters}
where $-j\le m\le j$.  The operators $J_\pm=J_1\pm{\rm i}J_2$,
$X_\pm=X_1\pm{\rm i}X_2$ and $X_3$ act on the vectors $|j,m;r\rangle$ as follows
\begin{mathletters}
\begin{eqnarray}
J_\pm |j,m;r\rangle &=& \sqrt{(j\mp m)(j\pm m+1)}\,|j,m\pm
1;r\rangle,\\
X_\pm |j,m;r\rangle
&=&\mp\frac{r\sqrt{(j\pm m+1)(j\pm m+2)}}
{\sqrt{(2j+1)(2j+3)}}|j+1,m\pm 1;r\rangle\nonumber\\
&&{}\pm\frac{r\sqrt{(j\mp m-1)(j\mp m)}}{\sqrt{(2j-1)(2j+1)}}
|j-1,m\pm 1;r\rangle,\\
X_3 |j,m;r\rangle
&=&\frac{r\sqrt{(j-m+1)(j+m+1)}}
{\sqrt{(2j+1)(2j+3)}}|j+1,m;r\rangle\nonumber\\
&&{}+\frac{r\sqrt{(j-m)(j+m)}}{\sqrt{(2j-1)(2j+1)}}|j-1,m;r\rangle.
\end{eqnarray}
\end{mathletters}
The orthogonality and completeness conditions satisfied by the
vectors $|j,m;r\rangle$ can be written as
\begin{eqnarray}
&&\langle j,m;r|j',m';r\rangle=\delta_{jj'}\delta_{mm'},\\
&&\sum_{j=0}^{\infty}\sum_{m=-j}^{j}
|j,m;r\rangle\langle j,m;r,|=I,
\end{eqnarray}
where $I$ is the identity operator.

We are now in a position to introduce the coherent states for a
particle on a sphere.  Namely, these states are defined as the
solution of the eigenvalue equation of the form
\begin{equation}
{\bf Z} |{\bf z}\rangle = {\bf z} |{\bf z}\rangle,
\end{equation}
where ${\bf Z}$ is given by
\begin{eqnarray}
{\bf Z} &=&\left(\frac{e^{\frac{1}{2}}}{\sqrt{1+4{\bf J}^2}}{\rm
sinh}\hbox{$\scriptstyle 1\over2 $}\sqrt{1+4{\bf
J}^2}+e^{\frac{1}{2}}{\rm cosh}\hbox{$\scriptstyle 1\over2 $}
\sqrt{1+4{\bf J}^2}\right){{\bf X}\over r}\nonumber\\
&&{}+{\rm i}\left(\frac{2e^{\frac{1}{2}}}{\sqrt{1+4{\bf J}^2}}{\rm sinh}
\hbox{$\scriptstyle 1\over2 $}\sqrt{1+4{\bf J}^2}\right){\bf
J}\times{{\bf X}\over r},
\end{eqnarray}
where the cross designates the vector product.  The operator ${\bf
Z}$ and ${\bf z}\in{\bf C}^3$ obey
\begin{equation}
{\bf Z}^2=1,\qquad {\bf z}^2=1.
\end{equation}
We also write down the following matrix representation of the
operator ${\bf Z}$ which is crucial for the algebraic analysis of
the problem:
\begin{equation}
e^{-(\bbox{\sigma}\bbox{\cdot}{\bf J}+1)}\bbox{\sigma}\bbox{\cdot}{\bf X}=
\bbox{\sigma}\bbox{\cdot}{\bf Z},
\end{equation}
where $\sigma_i$, $i=1,\,2,\,3$, are the Pauli matrices.

As with the standard coherent states we can generate the coherent
states from the ``fiducial vector'' $ |{\bf n}_3\rangle$ such that
\begin{equation}
{\bf Z} |{\bf n}_3\rangle={\bf n}_3 |{\bf n}_3\rangle,
\end{equation}
where ${\bf n}_3=(0,0,1)$, and
\begin{equation}
|{\bf
n}_3\rangle=\sum_{j=0}^{\infty}e^{-\frac{1}{2}j(j+1)}\sqrt{2j+1}|j,0;r\rangle.
\end{equation}
Namely, the coherent states are given by
\begin{equation}
|{\bf z}\rangle = \exp\left[\frac{{\rm arccosh}z_3}{\sqrt{1-z_3^2}}
({\bf z}\times{\bf n}_3)\bbox{\cdot}{\bf J}\right]
|{\bf n}_3\rangle.
\end{equation}
The projection of the coherent states (2.14) on the discrete basis
vectors $|j,m;r\rangle$ is
\begin{equation}
\langle j,m;r|{\bf z}\rangle = e^{-\frac{1}{2}j(j+1)}\sqrt{2j+1}\,
\frac{(2|m|)!}{|m|!}\sqrt{\frac{(j-|m|)!}{(j+|m|)!}}\left(
\frac{-\varepsilon(m)z_1+{\rm i}z_2}{2}\right)^{|m|} C_{j-|m|}^{|m|+\frac{1}{2}}
(z_3),
\end{equation}
where $\varepsilon(m)$ is the sign of $m$, and $C_n^\alpha(x)$ are
the Gegenbauer polynomials expressed with the help of the
hypergeometric function ${}_2F_1(a,b,c;z)$ by
\begin{equation}
C_n^\alpha(x) =\frac{\Gamma(n+2\alpha)}{\Gamma(n+1)\Gamma(2\alpha)}
\,{}_2F_1(-n,n+2\alpha,\alpha+\hbox{$\scriptstyle 1\over2 $};
\hbox{$\scriptstyle 1\over2 $}(1-x)).
\end{equation}
As mentioned earlier the coherent states are labelled by points of
the classical phase space $T^*S^2$.  The most natural complex
parametrization of the phase space discussed in \cite{1} is of the
form
\begin{equation}
{\bf z}=\cosh|{\bf l}|\,\frac{{\bf x}}{r}+{\rm i}\frac{\sinh|{\bf l}|}
{|{\bf l}|}\,{\bf l}\times \frac{{\bf x}}{r},
\end{equation}
where the vectors ${\bf l},\,{\bf x}\in{\bf R}^3$, fulfil
${\bf x}^2=r^2$ and ${\bf l}\bbox{\cdot}{\bf x}=0$, that is ${\bf l}$ is the
classical angular momentum and ${\bf x}$ is the radius vector of a particle
on a sphere.  Clearly, the vector ${\bf z}$ satisfies the second
equation of (2.10).
\section{Scalar product}
In this section we identify the Bargmann space of analytic functions
corresponding to the coherent states for a particle on a sphere
described above.  We now restrict, without loss of generality, to the case
with the unit sphere.  On introducing the spherical coordinates ${\bf
x}=(\sin\theta\cos\varphi,\, \sin\theta\sin\varphi,\, \cos\theta)$,
and parametrizing the tangent vector ${\bf l}$ by its norm $|{\bf
l}|\equiv l$ and the angle $\alpha$ between ${\bf l}$ and the
meridian passing through the point with the radius vector ${\bf x}$, we obtain
from (2.17) the following natural coordinates of the phase space
compatible with the constraints:
\begin{eqnarray}
z_1 &=& \cosh l\sin\theta \cos\varphi + {\rm i}\sinh l(\sin\alpha
\cos\varphi\cos\theta -\cos\alpha\sin\varphi),\nonumber\\
z_2 &=& \cosh l\sin\theta \sin\varphi + {\rm i}\sinh l(\sin\alpha
\sin\varphi\cos\theta+\cos\alpha\cos\varphi),\\
z_3 &=& \cosh l\cos\theta -{\rm i}\sinh
l\sin\alpha\sin\theta.\nonumber
\end{eqnarray}
Taking into account the fact that ${\bf z}$ transforms as the
vector we find that the Bargmann space should be specified by
\begin{equation}
\langle \phi|\psi\rangle = \frac{1}{8\pi^2}\int\limits_{0}^{2\pi}d\varphi
\int\limits_{0}^{\pi}\sin\theta d\theta
\int\limits_{0}^{2\pi}d\alpha \int\limits_{0}^{\infty}dl\,h(l)
(\phi({\bf z}^*(\theta,\varphi,\alpha,l))^*\psi({\bf z}^*(\theta,\varphi,
\alpha,l)),
\end{equation}
where $\phi ({\bf z}^*)=\langle {\bf z}|\phi\rangle$, ${\bf
z}^*=(z_1^*,z_2^*,z_3^*)$, $h(l)$ is an unknown density and
${\bf z}(\theta,\varphi,\alpha,l)$ is expressed by (3.1).
Clearly, the corresponding resolution of the identity can be written
in the form
\begin{equation}
\frac{1}{8\pi^2}\int\limits_{0}^{2\pi}d\varphi
\int\limits_{0}^{\pi}\sin\theta d\theta
\int\limits_{0}^{2\pi}d\alpha \int\limits_{0}^{\infty}h(l) dl\,
|{\bf z}(\theta,\varphi,\alpha,l)\rangle\langle{\bf z}(\theta,\varphi,
\alpha,l)| = I.
\end{equation}
In order to fix $h(l)$ consider the basis of the Bargmann space with
the scalar product (3.2)
\begin{equation}
e_{jm}({\bf z}(\theta,\varphi,\alpha,l))=\langle j,m|{\bf
z}\rangle,
\end{equation}
where $|j,m\rangle\equiv|j,m;1\rangle$ and $\langle j,m;r|{\bf z}\rangle$ is
given by (2.15) and (3.1).  Using (3.2) and (3.4) as well as some guessing
work we get
\begin{eqnarray}
&&\!\!\!\!\!\!\!\!\!\!\langle j,m|j',m'\rangle\nonumber\\
&=&\frac{1}{8\pi^2}\int\limits_{0}^{2\pi}d\varphi
\int\limits_{0}^{\pi}\sin\theta d\theta
\int\limits_{0}^{2\pi}d\alpha \int\limits_{0}^{\infty}dl\,h(l)
(e_{jm}({\bf z}^*(\theta,\varphi,\alpha,l))^*e_{j'm'}({\bf z}^*(\theta,\varphi,
\alpha,l))\nonumber\\
&=&
\delta_{jj'}\delta_{mm'}e^{-j(j+1)}\int\limits_{0}^{\infty}dl\,h(l)P_j(\cosh
2l),
\end{eqnarray}
where $P_n(z)$ are the Legendre polynomials such that
\begin{equation}
P_n(z)=\frac{1}{2^nn!}\frac{d^n}{dz^n}(z^2-1)^n.
\end{equation}
Thus the normalization condition for the orthonormal basis \{$ |j,m\rangle$\}
leads to the following equation on the density $h(l)$:
\begin{equation}
\int\limits_{0}^{\infty}dl\,h(l)P_j(\cosh 2l) = e^{j(j+1)}.
\end{equation}
We remark that the problem of the solution of this equation is
highly nontrivial (see acknowledgements) and it is related to the so 
called problem of moments \cite{4}.  We now recall that the Legendre 
polynomials satisfy the differential equation
\begin{equation}
\left((z^2-1)\frac{d^2}{dz^2}+2z\frac{d}{dz}\right)P_n(z)=n(n+1)P_n(z).
\end{equation}
From (3.8) it follows easily that
\begin{equation}
\frac{1}{\sinh \rho}\frac{d}{d\rho}\sinh \rho
\frac{d}{d\rho}P_n(\cosh \rho) = n(n+1)P_n(\cosh \rho ).
\end{equation}
We remark that the operator from the left hand side of (3.9) is
simply the Laplacian for the two-dimensional hyperbolic space.
Consider the heat kernel at the origin in hyperbolic space \cite{5},
given by
\begin{equation}
k_{H^2}(\rho,t) = 2^\frac{1}{2}(4\pi t)^{-\frac{3}{2}}e^{-\frac{t}{4}}
\int\limits_{\rho}^{\infty}\frac{se^{-\frac{s^2}{4t}}}{(\cosh s
-\cosh \rho)^\frac{1}{2}}ds.
\end{equation}
This heat kernel obeys the equation
\begin{equation}
\frac{\partial k_{H^2}}{\partial t} = \frac{1}{\sinh \rho}\frac{d}{d\rho}
\sinh \rho\frac{d}{d\rho}k_{H^2}(\rho,t),
\end{equation}
subject to the initial condition
\begin{equation}
2\pi \lim_{t\to0}\int\limits_{0}^{\infty}k_{H^2}(\rho,t)f(\rho)\sinh\rho
d\rho = f(0),
\end{equation}
where $f$ is an arbitrary continuous function with at most
exponential growth at infinity.  Putting $f(\rho)=P_n(\cosh\rho)$,
and making use of (3.11), (3.12) and the fact that $k_{H^2}(\rho,t)$ and
$\frac{d}{d\rho}k_{H^2}(\rho,t)$ decay faster-than-exponentially, we get
\begin{equation}
2\pi\int\limits_{0}^{\infty}k_{H^2}(\rho,t)P_n(\cosh\rho)\sinh\rho
d\rho = e^{tn(n+1)}.
\end{equation}
Hence, setting in (3.13) $\rho=2l$ and $t=1$, we finally find that
the desired density $h(l)$ satisfying (3.7) is
\begin{mathletters}
\begin{eqnarray}
h(l) &=& 4\pi k_{H^2}(2l,1)\sinh 2l\\
&=& \frac{e^{-\frac{1}{4}}\sinh 2l}{\sqrt{2\pi}}\int
\limits_{2l}^{\infty}\frac{se^{-\frac{s^2}{4}}}{(\cosh s
-\cosh 2l)^\frac{1}{2}}ds.
\end{eqnarray}
\end{mathletters}
We have thus identified the Bargmann space for the quantum mechanics
on a sphere specified by (3.2) and (3.14).  Taking into account
(3.1), (3.14a) and the relation which is an immediate consequence of
(2.17) such that
\begin{equation}
{\bf z}\bbox{\cdot}{\bf z}^* = |{\bf z}|^2 = \cosh 2l,
\end{equation}
the following form can be derived of the scalar product (3.2)
written with the help of the complex variables ${\bf z}$ (3.1) analogous
to the usual Bargmann representation \cite{6} for the standard coherent states:
\begin{equation}
\langle\phi|\psi\rangle =\int\limits_{{\bf z}^2=1}d\mu({\bf
z})(\phi({\bf z}^*))^*\psi({\bf z}^*),
\end{equation}
where
\begin{equation}
d\mu({\bf z}) = \hbox{$\scriptstyle 1\over4\pi $}k_{H^2}({\rm arccosh}
({\bf z}\bbox{\cdot}{\bf z}^*),1)dz_1dz_2dz_3dz_1^*dz_2^*dz_3^*,
\end{equation}
and $\phi({\bf z}^*)=\langle {\bf z}|\phi\rangle$.  Evidently, the
completeness of the coherent states can be written with the help of
the measure $d\mu({\bf z})$ as
\begin{equation}
\int\limits_{{\bf z}^2=1}d\mu({\bf
z}) |{\bf z}\rangle\langle{\bf z}| = I.
\end{equation}
\section{Reproducing kernel}
As is well-known the existence of the reproducing kernel is one of
the most characteristic properties of coherent states.  In view of
(3.18) the reproducing property can be written in the form
\begin{equation}
\phi({\bf w}^*) = \int\limits_{{\bf z}^2=1}d\mu({\bf z}){\cal K}({\bf
w}^*,{\bf z})\phi({\bf z}^*),
\end{equation}
where $\phi({\bf w}^*)=\langle {\bf w}|\phi \rangle$, and
\begin{equation}
{\cal K}({\bf w}^*,{\bf z}) = \langle {\bf w}|{\bf z} \rangle.
\end{equation}
It should be noted that the reproducing kernel ${\cal K}({\bf w}^*,
{\bf z})$ is the complex conjugate of the analytic function
\begin{equation}
\phi_{{\bf w}}({\bf z}^*) = \langle {\bf z}|{\bf w} \rangle
\end{equation}
representing the abstract coherent state $|{\bf w}\rangle$ also
called its symbol.  The formula on the overlap $\langle {\bf z}|{\bf w}
\rangle$ can be obtained from (2.14), (2.13) and (2.15).  Namely, we have
\begin{equation}
\langle {\bf z}|{\bf w} \rangle =
\sum_{j=0}^{\infty}e^{-j(j+1)}(2j+1)P_j({\bf z}^*\bbox{\cdot}{\bf w}),
\end{equation}
where $P_j(z)$ are the Legendre polynomials given by (3.6).
\section{Action of operators}
We now discuss the action of operators in the Bargmann representation.
We first observe that an immediate consequence of (2.8) is the
following formula on the action of operators ${\bf Z}^\dagger$:
\begin{equation}
{\bf Z}^\dagger \phi({\bf z}^*) = {\bf z}^*\phi({\bf z}^*),
\end{equation}
where $\phi({\bf z}^*)=\langle {\bf z}|\phi\rangle$ and we recall that
${\bf z}^2=1$.  Now consider the action of the operator ${\bf J}^2$.
By (2.4a) the action of the operator ${\bf J}^2$ on the basis
$e_{jm}({\bf z}^*)=\langle{\bf z}|j,m\rangle$ of the Bargmann space
is the following one:
\begin{equation}
{\bf J}^2 e_{jm}({\bf z}^*) = j(j+1)e_{jm}({\bf z}^*).
\end{equation}
Using (2.15), the differential equation satisfied by the Gegenbauer
polynomials of the form
\begin{equation}
\left((z^2-1)\frac{d^2}{dz^2}+(2\lambda+1)\frac{d}{dz}-n(2\lambda+n)\right)
C_n^\lambda(z)=0,
\end{equation}
and (5.2) we find that the operator ${\bf J}^2$ acts in the
representation (3.16) as follows
\begin{equation}
{\bf J}^2\phi({\bf z}^*) = -\left({\bf
z}^*\times\frac{\partial}{\partial{\bf z}^*}\right)^2\phi({\bf z}^*).
\end{equation}
Taking into account (5.4) and (5.1) we obtain
\begin{equation}
{\bf J}\phi({\bf z}^*) = -{\rm i}\left({\bf z}^*\times\frac{\partial}
{\partial{\bf z}^*}\right)\phi({\bf z}^*).
\end{equation}
The relation (5.5) can be easily checked on the basis $e_{jm}({\bf
z}^*)$ with the help of (2.15), (2.4a) and (2.5a).  Further, using
(2.5c), (2.15) and elementary properties of the Gegenbauer
polynomials we get
\begin{equation}
X_3\phi({\bf z}^*) = e^{-\frac{1}{2}{\bf J}^2}z_3^*
e^{\frac{1}{2}{\bf J}^2}\phi({\bf z}^*),
\end{equation}
where the action of ${\bf J}^2$ is given by (5.4).
The action of the remaining coordinates of the position operator
${\bf X}$ can be obtained by means of the following identity
describing the complex rotation of ${\bf X}$:
\begin{equation}
e^{{\bf w}\bbox{\cdot}{\bf J}}{\bf X}e^{-{\bf w}\bbox{\cdot}{\bf J}}=
\cosh\sqrt{{\bf w}^2}\,{\bf X}-{\rm i}\frac{\sinh\sqrt{{\bf w}^2}}
{\sqrt{{\bf w}^2}}
{\bf w}\times{\bf X}+\frac{1-\cosh\sqrt{{\bf w}^2}}{{\bf w}^2}{\bf w}
({\bf w}\bbox{\cdot}{\bf X}).
\end{equation}
Namely, we have
\begin{mathletters}
\begin{eqnarray}
X_1\phi({\bf z}^*) &=& -\frac{{\rm i}}{\sinh1}\left(e^{J_2-\frac{1}{2}
{\bf J}^2}z_3^*e^{\frac{1}{2}{\bf J}^2-J_2}-\cosh1 e^{-\frac{1}{2}{\bf J}^2}
z_3^*e^{\frac{1}{2}{\bf J}^2}\right)\phi({\bf z}^*),\\
X_2\phi({\bf z}^*) &=& \frac{{\rm i}}{\sinh1}\left(e^{J_1-\frac{1}{2}
{\bf J}^2}z_3^*e^{\frac{1}{2}{\bf J}^2-J_1}-\cosh1 e^{-\frac{1}{2}{\bf J}^2}
z_3^*e^{\frac{1}{2}{\bf J}^2}\right)\phi({\bf z}^*),
\end{eqnarray}
\end{mathletters}
where the action of the operators $J_i$, $i=1,\,2$, and ${\bf J}^2$
is given by (5.5) and (5.4), respectively.
Finally, taking into account the identity
\begin{equation}
{\bf Z} = e^{-\frac{1}{2}{\bf J}^2}{\bf X}e^{\frac{1}{2}{\bf J}^2},
\end{equation}
which is a straightforward consequence of (2.11) and the commutation
relation
\begin{equation}
[{\bf J}^2,\bbox{\sigma}\bbox{\cdot}{\bf X}] = -2(\bbox{\sigma}\bbox{\cdot}{\bf
J}+1)\bbox{\sigma}\bbox{\cdot}{\bf X},
\end{equation}
following directly from (2.1) and (2.3), we obtain the action of the
operator ${\bf Z}$.  It follows that
\begin{mathletters}
\begin{eqnarray}
Z_1\phi({\bf z}^*) &=& -\frac{{\rm i}}{\sinh1}\left(e^{J_2-
{\bf J}^2}z_3^*e^{{\bf J}^2-J_2}-\cosh1 e^{-{\bf J}^2}
z_3^*e^{{\bf J}^2}\right)\phi({\bf z}^*),\\
Z_2\phi({\bf z}^*) &=& \frac{{\rm i}}{\sinh1}\left(e^{J_1-
{\bf J}^2}z_3^*e^{{\bf J}^2-J_1}-\cosh1 e^{-{\bf J}^2}
z_3^*e^{{\bf J}^2}\right)\phi({\bf z}^*),\\
Z_3\phi({\bf z}^*) &=& e^{-{\bf J}^2}z_3^*
e^{{\bf J}^2}\phi({\bf z}^*).
\end{eqnarray}
\end{mathletters}
\section{The Bargmann representation and the coordinate representation}
In this section we discuss the relationship between the introduced
Bargmann representation and the standard coordinate representation
for the quantum mechanics on a sphere.  We begin with recalling the
basic facts about the coordinate representation.  Consider the
position operators ${\bf X}$ for a particle on a sphere satisfying
the $e(3)$ algebra (2.1).  Recall that we restrict to the
irreducible representations which fulfil (2.3) and ${\bf X}^2=1$.
The coordinate representation is spanned by the common eigenvectors
$ |{\bf x}\rangle$ of the position operators such that
\begin{equation}
{\bf X}|{\bf x}\rangle = {\bf x} |{\bf x}\rangle,
\end{equation}
where ${\bf x}^2=1$.  The resolution of the identity is of the form
\begin{equation}
\int\limits_{{\bf x}^2=1}d\nu({\bf
x}) |{\bf x}\rangle\langle{\bf x}| = I,
\end{equation}
where $d\nu({\bf x})=d\nu(\theta,\varphi)=\sin\theta d\varphi d\theta$,
accordingly to the natural i.e.\ spherical coordinates ${\bf x}=
(\sin\theta\cos\varphi,\,\sin\theta\sin\varphi,\, \cos\theta)$
compatible with the constraint ${\bf x}^2=1$.  The
completeness gives rise to a functional representation of vectors
such that
\begin{equation}
\langle\phi|\psi\rangle =\int\limits_{{\bf x}^2=1}d\nu({\bf
x})\phi^*({\bf x})\psi({\bf x}),
\end{equation}
where $\phi({\bf x})=\langle {\bf x}|\phi\rangle$.  Clearly, we can write
the completeness condition (6.2) and the scalar product (6.3) as
\begin{equation}
\int\limits_{0}^{2\pi}d\varphi\int\limits_{0}^{\pi}\sin\theta d\theta
|\theta,\varphi\rangle\langle \theta,\varphi| = I,
\end{equation}
where $|\theta,\varphi\rangle\equiv |{\bf x}\rangle$, and
\begin{equation}
\langle\phi|\psi\rangle = \int\limits_{0}^{2\pi}d\varphi\int\limits_{0}^{\pi}
\sin\theta d\theta\, \phi^*(\theta,\varphi)\psi(\theta,\varphi),
\end{equation}
where $\phi(\theta,\varphi)=\langle \theta,\varphi|\phi\rangle$,
respectively.  The passage from the coordinate representation to the
angular momentum representation generated by the vectors $|j,m\rangle$
satisfying (2.4) with $r=1$ is given by
\begin{equation}
\langle \theta,\varphi|j,m\rangle =
Y_{jm}(\theta,\varphi)=(-1)^{\frac{m-|m|}{2}}\sqrt{\frac{(2j+1)(j-|m|)!}
{4\pi(j+|m|)!}}P_j^{|m|}(\cos\theta)e^{{\rm i}m\varphi},
\end{equation}
where $Y_{jm}(\theta,\varphi)$ are the spherical harmonics and
$P_n^m(z)$ are the associated Legendre polynomials which can be
defined by
\begin{equation}
P_n^m(z) = (-1)^m(1-z^2)^{\frac{m}{2}}\frac{d^m}{dz^m}P_n(z),
\end{equation}
where $P_n(z)$ are the Legendre polynomials given by (3.6).  Of
course, $Y_{jm}(\theta,\varphi)$ form the orthonormal basis of the
Hilbert space of the square integrable functions on a sphere $S^2$
specified by the scalar product (6.5).  Taking into account (6.6)
and the identity
\begin{equation}
C_{n-m}^{m+\frac{1}{2}}(z) = (-1)^m\frac{(1-z^2)^{-\frac{m}{2}}m!2^m}
{(2m)!}P_n^m(z),
\end{equation}
where $m+1$ is natural, we find that the kernel (6.6) can be written
in the form analogous to (2.15) such that
\begin{equation}
\langle {\bf x}|j,m\rangle = \sqrt{\frac{2j+1}{4\pi}}\,
\frac{(2|m|)!}{|m|!}\sqrt{\frac{(j-|m|)!}{(j+|m|)!}}\left(
\frac{-\varepsilon(m)x_1-{\rm i}x_2}{2}\right)^{|m|} C_{j-|m|}^{|m|+
\frac{1}{2}}(x_3).
\end{equation}
Now, let $|({\bf x},{\bf l})\rangle$ designate the coherent state
$|{\bf z}\rangle$ in accordance with the parametrization of the
phase space given by (2.17).  Eqs.\ (5.9), (2.15) and (6.9) taken
together yield
\begin{equation}
|({\bf x},{\bf 0})\rangle = \sqrt{4\pi}e^{-\frac{1}{2}{\bf J}^2}
|{\bf x}\rangle.
\end{equation}
Using (2.14), (6.10) and (2.13) and proceeding as with (4.4) we find
that the passage from the coordinate representation to the coherent
states representation is described by the matrix element
\begin{equation}
\langle {\bf x}|{\bf z} \rangle = \frac{1}{\sqrt{4\pi}}
\sum_{j=0}^{\infty}e^{-\frac{1}{2}j(j+1)}(2j+1)P_j({\bf x}\bbox{\cdot}{\bf z}).
\end{equation}
Evidently, (6.11) defines a unitary map $U:\phi\to\tilde\phi$ from
the standard Hilbert space of square integrable functions on the
sphere $S^2$ with the scalar product (6.3) onto the Bargmann space
of analytic functions specified by the scalar product (3.16), of the
form
\begin{equation}
(U\phi)({\bf z}^*) = \int\limits_{{\bf x}^2=1}d\nu({\bf
x})k({\bf x}\bbox{\cdot}{\bf z}^*)\phi({\bf x}),
\end{equation}
where $k({\bf x}\bbox{\cdot}{\bf z})=\langle {\bf x}|{\bf z} \rangle$.  The
inverse operator $U^{-1}$ is given by
\begin{equation}
(U^{-1}\tilde\phi)({\bf x}) = \int\limits_{{\bf z}^2=1}d\mu({\bf
z})k({\bf x}\bbox{\cdot}{\bf z})\tilde\phi({\bf z}^*).
\end{equation}

We finally discuss the probability density $p_{{\bf z}}({\bf x})$
for the coordinates in the normalized coherent state $|{\bf z}\rangle
/\sqrt{\langle {\bf z}|{\bf z}\rangle}$ such that
\begin{equation}
p_{{\bf z}}({\bf x}) = \frac{|\langle {\bf x}|{\bf z} \rangle|^2}
{\langle {\bf z}|{\bf z}\rangle}.
\end{equation}
We recall that (6.14) is also called, especially in the context of
the theory of quantum chaos the Husimi representation for the
localized state on the sphere $|{\bf x}\rangle$.  Let ${\bf
z}=\cosh|{\bf l}|\overline{{\bf x}}+{\rm i}\frac{\sinh|{\bf l}|}
{|{\bf l}|}\,{\bf l}\times \overline{{\bf x}}$ (see (2.17)), so
$\overline{{\bf x}}$ corresponds to the position and ${\bf l}$ to
the angular momentum of a particle on a sphere.  From computer
simulations it follows that for small enough $|{\bf l}|$ the
function $p_{{\bf z}}({\bf x})$ is peaked at ${\bf x}=\overline{{\bf x}}$.
Therefore the parameter ${\bf x}$ in the formula (2.17) can be really
regarded as the classical position for a particle on a sphere.
\section{Discussion}
In this work we have introduced the Bargmann representation
referring to the coherent states for a particle on a sphere.   The
very general construction of the Bargmann space, where the
configuration space is a symmetric space has been recently
introduced by Stenzel \cite{7}.  As remarked by Hall \cite{8} such
construction generalizes the case discussed herein with the
configuration space coinciding with the sphere $S^2$.  Nevertheless, it
does not fit into the usual scheme of construction of Bargmann spaces by
means of the resolution of the identity for the corresponding coherent states.
More precisely, the coherent states are not utilized at all in
\cite{7}.  The approach taken up in \cite{7} is very general and as
far as we are aware the observations of our work are one of the
first concrete nontrivial example of the general construction
discussed in \cite{7}.  On the other hand, the construction
introduced by Stenzel shows that the formalism introduced in this
paper has a deeper mathematical context.  We finally point out that 
the results obtained herein seem to be of interest also in the theory of
classical orthogonal polynomials in the complex domain as well
as the theory of heat kernels.
\section*{Acknowledgements}
We would like to thank Brian Hall for invaluable comments,
especially for pointing out the solution to equation (3.7) in
terms of the heat kernel for the hyperbolic space.  We are convinced
that without the solution of (3.7) provided by Brian we could not
write this paper.


\begin{references}
\bibitem{1}K. Kowalski and J. Rembieli\'nski, J. Phys. A: Math. Gen. 
{\bf 33}, 6035 (2000).
\bibitem{2}R. J. Glauber, Phys. Rev {\bf 130}, 2529 (1963); {\bf
131}, 2766 (1963); J.R. Klauder, J. Math. Phys. {\bf 4}, 1055 (1963).
\bibitem{3}J.R. Klauder and B.S. Skagerstam, {\em Coherent
States--Applications in Physics and Mathematical Physics} (World
Scientific, Singapore, 1985).
\bibitem{4}J.A. Shohat and J.D. Tamarkin, {\em The Problem of Moments},
Mathematical Surveys, No 1, Providence, R.I. Am. Math. Soc., 1950.
\bibitem{5}E.B. Davies, {\em Heat Kernels and Spectral Theory}
(Cambridge Univ. Press, Cambridge, 1989); 
R. Camporesi, {\em Harmonic Analysis and Propagators on Homogeneous
Spaces}, Phys. Rep. {\bf 196}, 1--134 (1990).
\bibitem{6}V. Bargmann, Commun. Pure Appl. Math. {\bf 14}, 187 (1961).
\bibitem{7}M.B. Stenzel, J. Funct. Anal. {\bf 165}, 44 (1999); B.C.
Hall, J. Funct. Anal. {\bf 122}, 103 (1994).
\bibitem{8}B.C. Hall, Preprint arXiv: quant-ph/0006037 (2000).
\end{references}
\end{document}